\documentclass[fleqn,10pt]{wlscirep}
\usepackage{dcolumn}
\usepackage{amssymb,amsmath}
\usepackage{color}
\usepackage{epsfig}

\newcommand{\blue}{\textcolor{black}}

\title{Micro-thermocouple on nano-membrane: thermometer for nanoscale measurements}

\author[1,2,+]{Armandas Bal\v{c}ytis}
\author[3,+]{Meguya Ryu}
\author[1,4]{Saulius Juodkazis}
\author[3]{Junko Morikawa}
\affil[1]{Swinburne University of Technology, John st., Hawthorn, 3122 Vic, Australia}
\affil[2]{Center for Physical Sciences and Technology, Savanoriu ave. 231, LT-02300 Vilnius, Lithuania}
\affil[3]{Tokyo Institute of Technology, Meguro-ku, Tokyo 152-8550, Japan}
\affil[4]{Melbourne Center for Nanofabrication, Australian National Fabrication Facility, Clayton 3168 Vic, Melbourne, Australia}

\affil[+]{these authors contributed equally to this work}
\begin{abstract}
A thermocouple of Au-Ni with only 2.5-$\mu$m-wide electrodes on a 30-nm-thick Si$_3$N$_4$ membrane was fabricated by a simple low-resolution electron beam lithography and lift off procedure. The thermocouple is shown to be sensitive to heat generated by laser as well as an electron beam. \blue{N}ano-thin membrane \blue{was used to reach a high} spatial resolution of energy deposition and \blue{to} realise a heat source of \blue{sub-}$1~\mu$m diameter. \blue{This was achieved due to a limited generation of secondary electrons, which increase a lateral energy deposition.} A low thermal capacitance of the fabricated devices is useful for \blue{the} real time  monitoring of small \blue{and fast} temperature changes, e.g., due to convection, and can be detected through an optical and mechanical barrier of the nano-thin membrane. Temperature changes up to $\sim 2\times 10^5$~K/s can be measured at 10~kHz rate. A simultaneous down-sizing of both, the heat detector and heat source strongly required for creation of thermal microscopy is demonstrated. \blue{Peculiarities} of Seebeck constant (thermopower) dependence on electron injection into thermocouple are discussed. \blue{Modeling of thermal flows on a nano-membrane with presence of a micro-thermocouple was carried out to compare with experimentally measured temporal response.}
 
\end{abstract}
\begin{document}

\flushbottom
\maketitle
% * <john.hammersley@gmail.com> 2015-02-09T12:07:31.197Z:
%
%  Click the title above to edit the author information and abstract
%
\thispagestyle{empty}
\section*{Introduction}

%_________________________Fig. 1
\begin{figure}[b]
\centering
\includegraphics[width=1\linewidth]{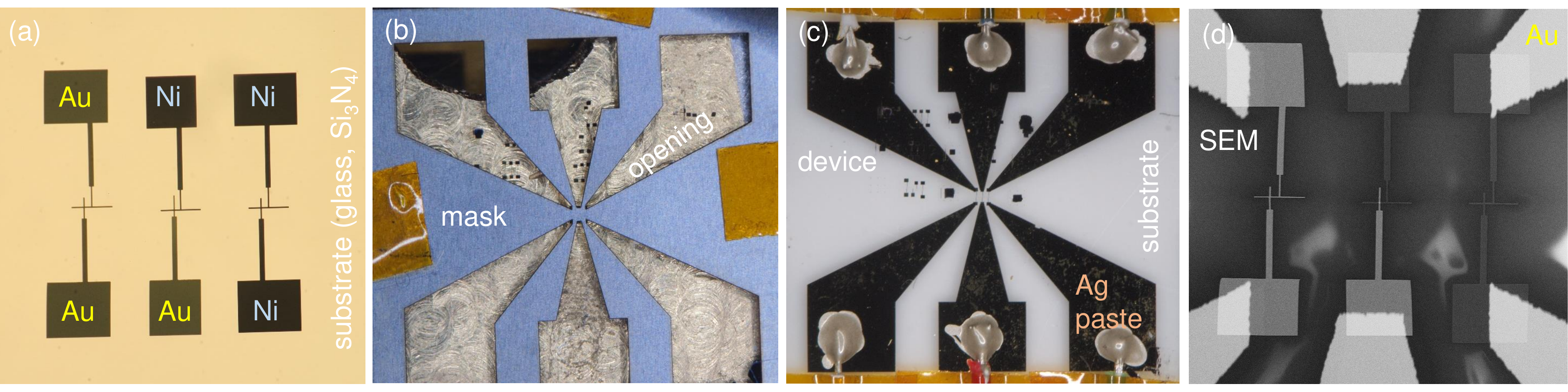}
\caption{Thermocouple fabrication stages and patterns of electrodes at different magnifications. Thermocouples were fabricated on glass and Si$_3$N$_4$ membranes of different thicknesses: 1~$\mu$m and 30~nm. (a) Metal junctions of 2.5-$\mu$m-wide metal stripes with $100\times 100~\mu$m$^2$ primary contact pads. (b) Photo image of a laser ablated photolithography mask \blue{used} for resist exposure. It defines the secondary contact pads to interface with electrical measurements. (c) Photo of the final device on glass. (d) \blue{A} SEM image of the micro-thermocouple \blue{and reference electrodes}. The central pair is \blue{the} Au-Ni thermocouple.}
\label{f-samp}
\end{figure}
%_________________________Fig. 2
\begin{figure}[ht]
\centering
\includegraphics[width=0.85\linewidth]{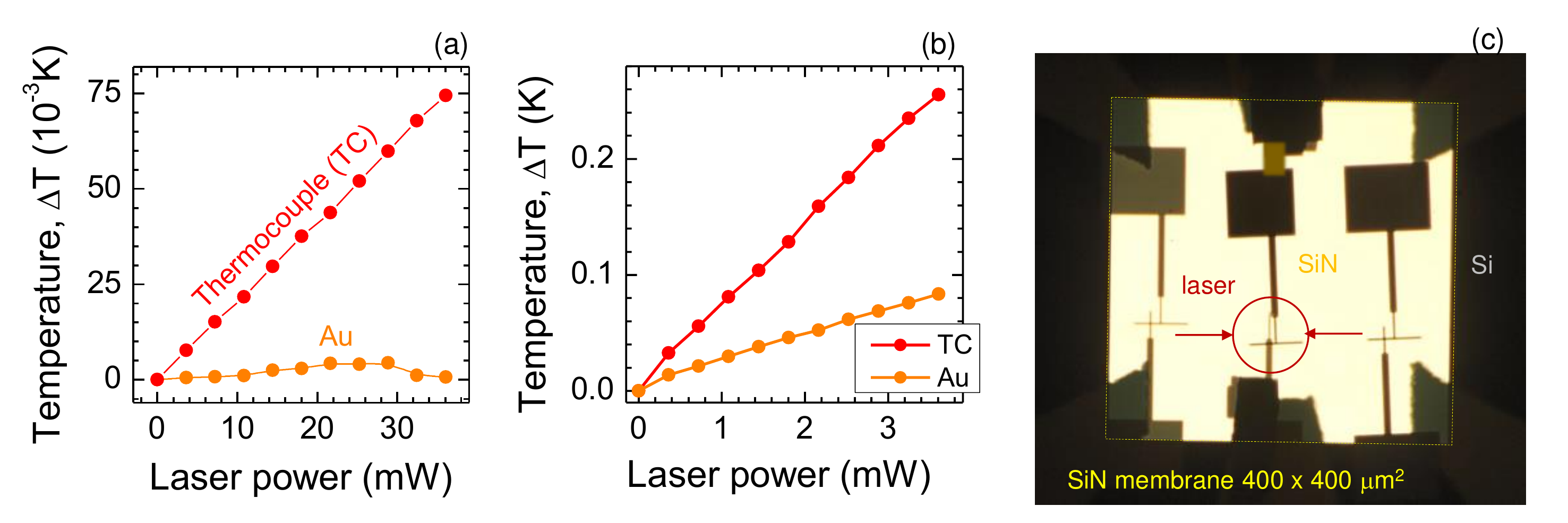}
\caption{Characterisation of thermocouples. (a) Temperature increase induced by laser heating at different laser power\blue{s} measured by the \blue{optical modulation} method (wavelength $\lambda = 830$~nm; $p$-polarisation at slanted front-side incidence). Thermocouple Au-Ni was made on a slide glass. \blue{Sensitivity of} $10.1~\mu$V/K~\cite{Chu1} \blue{determined for similar thermocouple was used for estimation of temperature changes}; Au-Au junction was used as a reference. Illumination of the substrate was \blue{carried out} from the side to contacts. (b) Temperature vs. laser power for 1~$\mu$m-thick Si$_3$N$_4$ membrane. \blue{The Au-Au reference electrodes had a Cr adhesion layer and formed a thermocouple which was experiencing a thermal gradient due to asymmetry of the primary contact pads during laser heating (see panel (c)).} (c) An optical see-through image of the $400\times 400~\mu$m$^2$ SiN-membrane region with thermocouple whose response is plotted in (b); laser spot was $\sim 100~\mu$m in diameter. \blue{Note a thermal asymmetry of this layout where the upper $100\times 100$~$\mu$m$^2$ primary contact square pad was on the SiN membrane while the lower one on the Si substrate.}}
\label{f-thick}
\end{figure}

Thermal characterisation of nanoscale heat sources and heat flows around/through nano-objects is a challenging task~\cite{Chae} due to a deep sub-wavelength nature when IR imaging is used while a direct contact measurement suffers from a large heat capacitance and, correspondingly, alters thermal distribution pattern. Moreover, direct contact methods \blue{of temperature measurements} are slow when micro-thermocouples are used~\cite{10oe8300}. Simulation of \blue{a} heat transport by atomistic methods, e.g., Monte Carlo simulations, still have challenges for modeling of the actual sizes of nanoscale devices~\cite{Cahill}. During the last decade advances in measuring heat transport through the interfaces, at conditions of phase transitions with nanoscale resolution using an atomic force microscopy (AFM) probe\blue{s} were reported~\cite{Cahill1}. A real time monitoring capability is still strongly required for research of phase transitions, crystalline phase formations and photo-thermal cancer treatments~\cite{Neal}. For \blue{an} optical light harvesting, the thermal radiation and suppression of reflectivity (impedance matching~\cite{Mazilu}) have to be determined for the optimised \blue{performance}. 

A recently introduced hot-tip scanning lithography \blue{with an} AFM tip heated up to $\sim 800^\circ$C temperature (Nanofrazor, SwissLitho, Ltd.) allows to write 3D nanoscale patterns with resolution down to 10~nm in molecular glass resists. With this approach, a secondary electron damage usually occurring in a high-resolution electron beam lithography (EBL) exposure \blue{during patterning of thin layers of electronic devices} is avoided. Thermal protocols of 3D material growth and structuring for nanotechnology applications (\blue{a} recent review~\cite{17n923}) are strongly dependent on thermal properties and conditions, which are currently not well known at the nanos\blue{c}ale. Management of temperature and heat flows in 2D layered \blue{materials and} structures, \blue{e.g.,} graphene, are important for photo-detectors and light harvesting \blue{devices}~\cite{Frank}. Conceptually, a thermal microscopy with a miniaturised heat source and detector are strongly required to develop next generation of transistors beyond current 10-nm-node where thermal management will be of paramount importance.   

We show here fabrication and characterisation of thermocouples on 30-nm-thick Si$_3$N$_4$ membranes. \blue{The} Au-Ni thermocouple \blue{was} made \blue{from} thin evaporated metal films of $\sim 100$~nm thicknesses. \blue{S}mall thermal capacitance of SiN nano-membrane \blue{facilitated detection of} minute temperature changes due to the absorbed energy (dose), e.g., $\sim 0.1$~K measured under $\sim 1$~mW red laser illumination as well as heating by an electron beam exposure \blue{(this estimate was obtained for the sensitivity of 10.1~$\mu$V/K determined for a similar thermocouple~\cite{Chu1})}.   

%_________________________Fig. 3
\begin{figure}[ht]
\centering
\includegraphics[width=1\linewidth]{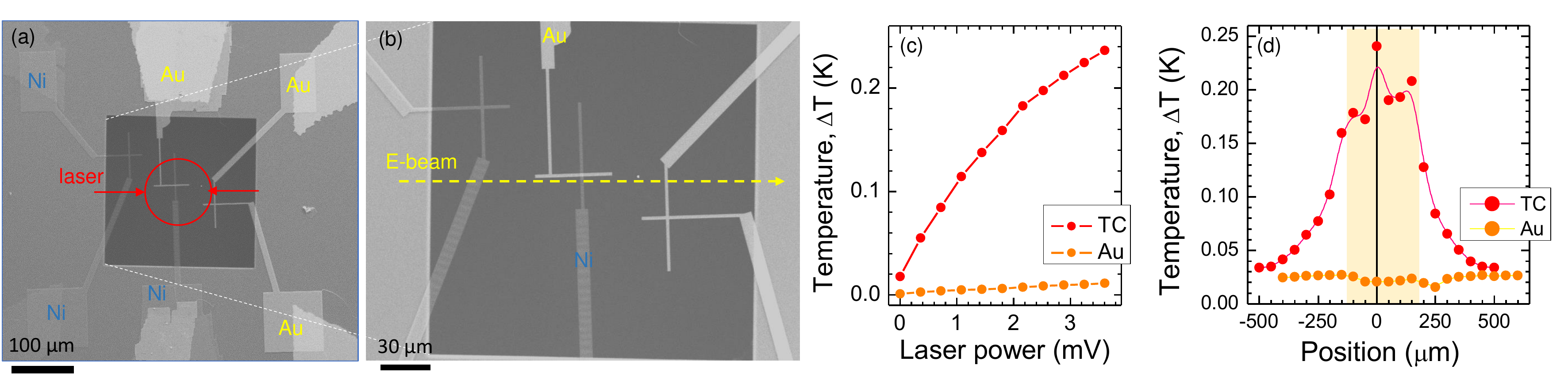}
\caption{Thermocouple on a 30-nm-thick SiN-membrane. (a,b) SEM images of thermocouple made \blue{on a} $250\times 250~\mu$m$^2$ SiN window. Note, secondary (large) contact pads are made from the same metal (Au or Ni) as the smaller ones to avoid formation of a secondary thermocouple. The large contacts are placed on Si substrate to \blue{remove a} thermal gradient on the thermocouple \blue{(such gradient was responsible for the observed temperature change in Fig.~\ref{f-thick}(b) measured with the Au-Au contact)}. (c) Temperature vs. laser power. (d) Temperature vs. \blue{position of the} electron beam across the central cross section (along the line in (b)). Diameter of the e-beam \blue{was} $\sim 0.5~\mu$m at the acceleration voltage of 25~keV; modulation frequency was 30~Hz, current $\sim 1.7$~nA as measured by Faraday cup without \blue{the} sample. Central shaded region depicts \blue{the} location of membrane. E-beam was scanned across the SiN window and \blue{Si} substrate without \blue{direct} exposure of metal leads/contacts.}
\label{f-thin}
\end{figure}

\section*{Results}\label{Res}

\subsection{\blue{Laser beam heating}}

Miniaturisation of \blue{a} thermocouple by \blue{fabrication of a} cross pattern of few-micrometer-wide stripes of dissimilar metals is an obvious step~\cite{10oe8300} and was demonstrated for direct contact measurements of \blue{the} temperature diffusivity in polymers~\cite{09apa551}. \blue{The n}ext improvement carried out in this study was \blue{fabrication of} such pattern onto a thin Si$_3$N$_4$ membrane to reduce thermal capacitance and augment sensitivity as well as \blue{to} reduce a response time of thermocouple. Figure~\ref{f-samp} shows the \blue{pattern of} thermocouple used in this study, a mask for definition of contacts pads, the final device, and \blue{a} scanning electron microscopy (SEM) image of the Au-Ni micro-thermocouple \blue{made} on a slide glass substrate. 

Voltage generated by the thermocouple in response \blue{to} modulated laser power when laser was illuminated from the front-side (the surface on which the contacts were made) is shown in Fig.~\ref{f-thick}(a). Sensitivity increased $\sim 34.5$ times for the same thermocouple made on a 1-$\mu$m-thick SiN-membrane (Fig.~\ref{f-thick}(b)). \blue{For temperature calibration, the sensitivity $10.1~\mu$V/K of a similar thermocouple was used~\cite{Chu1}. In this study we were aiming at  detection of fast temperature changes induced by the laser and electron beam irradiation rather on determination of its absolute values (see, Methods Sec. for details).}
A junction of Au-Au used as a reference also showed some sensitivity for the laser-induced heating. This \blue{is caused by two reasons. First, an adhesion of 5-to-10-nm-thick layer of Cr was evaporated before deposition of Au. This created an additional Au-Cr thermocouple. The second reason is due to an asymmetry in placement of the square} $100\times 100~\mu$m$^2$ contact pads, \blue{which have the bi-metal Cr-Au structure (the pads are seen in upper and lower part in Fig.~\ref{f-thick}(c)). The upper pad was placed on the SiN membrane while the lower one was in contact with the Si substrate. This caused an unwanted temperature gradient upon laser heating and the temperature sensitivity. In next design the both pads were placed on Si to eliminate temperature gradient and sensitivity of the Au-Au reference contacts.}

In the final design, a 30-nm-thick SiN-membrane with a smaller window was used. It secured placement of the contact pads outside the membrane region, hence, at a constant temperature defined by the bulky Si substrate. Figures~\ref{f-thin}(a,b) show SEM images of the thermocouple. Response of the thermocouple to laser heating was similar as for 1~$\mu$m membrane. Saturation tendency at a larger laser power is attributable to the heat sink effect of the substrate, which was closer to the $\sim 100~\mu$m diameter laser spot  (note, the smaller membrane window). \blue{The} laser irradiation at a slanted angle caused a larger elliptical projection of the laser spot onto the membrane. The Au-Au junction showed no photo-sensitivity when the contact pads were outside the membrane region in this final thermocouple design. 

\subsection{\blue{Electron beam irradiation}}

Thermal sensitivity of the thermocouple to electron beam focused to the $\sim 0.5~\mu$m-diameter spot and scanned across the membrane with single point irradiation at 30~Hz is shown in Fig.~\ref{f-thin}(d). The largest voltage response was recorded with e-beam close to the thermocouple. Here we used the same 10.1~$\mu$V/K coefficient to estimate $\Delta T$ and validity of this judgment is discussed below. When separation \blue{between} the e-beam spot \blue{and thermocouple was} $>5~\mu$m \blue{with e-beam} still on the SiN-membrane, the temperature readout was almost the same. When e-beam was on \blue{the} thick Si substrate, thermocouple was recording a decreasingly smaller temperature as electrons were impinging at \blue{a} larger distance from the thermocouple. The slope of the voltage with distance had a characteristic single exponential decay over distance $x_d = 130~\mu$m. With e-beam directly irradiating the Au-Ni junction or the metal leads (Au or Ni) there was an electrical signal generated due to electron injection and was by two orders of magnitude larger ($\sim 230~\mu$V vs $\sim 2~\mu$V for the electron and laser exposures, respectively). All the e-beam exposure locations were selected to \blue{avoid direct electron irradiation of the} metal leads.    
 
 %_________________________Fig. 4
\begin{figure}[ht]
\centering
\includegraphics[width=0.8\linewidth]{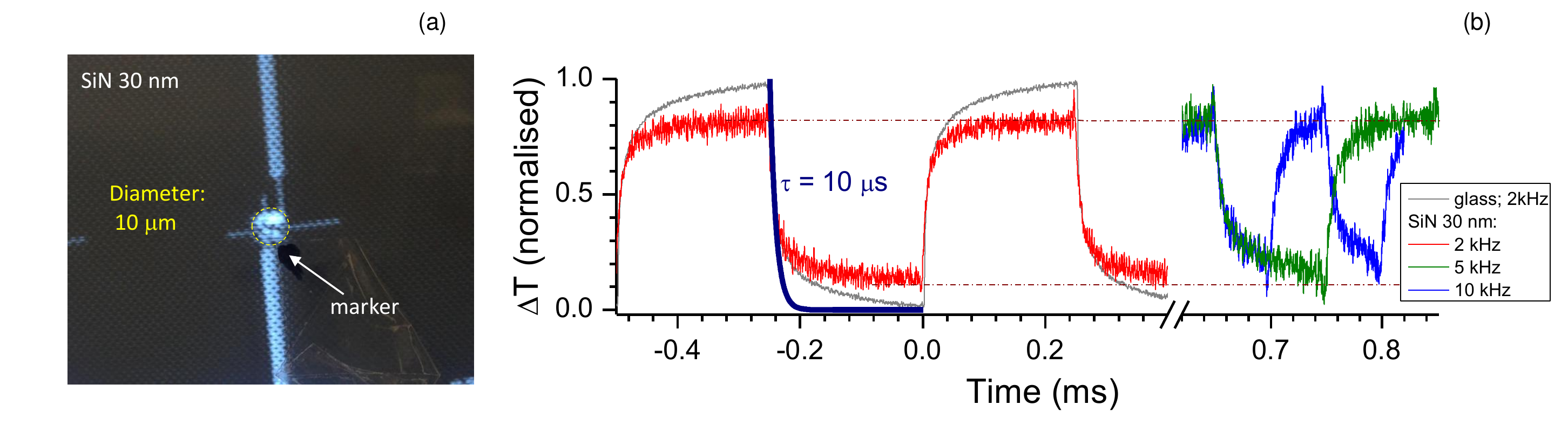}
\caption{Temporal response of thermocouple on a 30-nm-thick SiN-membrane to a square-wave optical excitation. (a) Video image of a tightly-focused laser beam onto thermocouple with $\sim 10~\mu$m spot diameter; $\lambda = 830$~nm. (b) Temporal response of thermocouples: (\emph{i}) to a 3.6~mW laser power at repetition rate $f = 2$~kHz with a thermocouple on a slide glass and (\emph{ii}) to 1.8~mW power with thermocouple on \blue{a} 30-nm-thick SiN-membrane at $f = 2, 5, 10$~kHz; note different x-axis scales in (b). The fastest switching time was $\tau= RC = 10~\mu$s  with ohmic resistance of thermocouple $R = 500~\Omega$ and $C = 20$~nF. Electronic pre-amplifier of 100$^\times$ was used for a direct observation by oscilloscope.  The \blue{estimated} max-min $\Delta T$ span was 16~K (30.7-to-14.7~K above RT of 22$^\circ$C) for the thermocouple on glass and $\Delta T = 4.1$~K for 30~nm SiN-membrane (31.3-to-27.2~K) at 2~kHz; at higher 5 and 10~kHz frequencies the max temperature increase was similar $\Delta T_{max} = 30.5$~K and min-max span of $\sim$4~K. }
\label{f-fast}
\end{figure}

\subsection{\blue{Detection of fast heat transients}}

Next, temporal response of thermocouples with $6~\mu$m$^2$ area were investigated at much higher $\sim$kHz frequencies with more tight definition of the  laser focal spot on the junction (Fig.~\ref{f-fast}). A square-wave excitation was used to test thermocouple response time. Absorbed amount of light irradiated from the contact side is very small \blue{in the case of 30-nm-thick SiN }, however, the saturation level of signal is reached faster as compared with the same thermocouple on the slide glass. The fastest segment of temperature rise and decay was $\tau\sim 10~\mu$s (the same time constant was the best fit also for the rising part of the transient but is not shown in Fig.~\ref{f-fast}). Up to \blue{a} 10~kHz laser repetition rate, saturated temperature values were \blue{reached} within the period of illumination. The absolute temperature values were extracted from oscillograms (Fig.~\ref{f-fast}). For the 30~nm SiN-membrane, an approximately 4-times smaller span of min-max temperatures was observed as compared with thermocouple on the slide-glass (4 vs 16~K). Also, slightly higher maximum temperatures were reached on the SiN-membrane. Considering a half of the min-max span of $\Delta T\approx 2$K occurring within the fastest change of $\tau=10~\mu$s, the heating (cooling) rates up to $0.2\times 10^6$~K/s are measurable. This is a promising feature for practical applications in real time monitoring of temperature. 
 
\subsection{\blue{Heat and direct electron injection contributions}} 
 
\begin{figure}[h]
\centering
\includegraphics[width=0.85\linewidth]{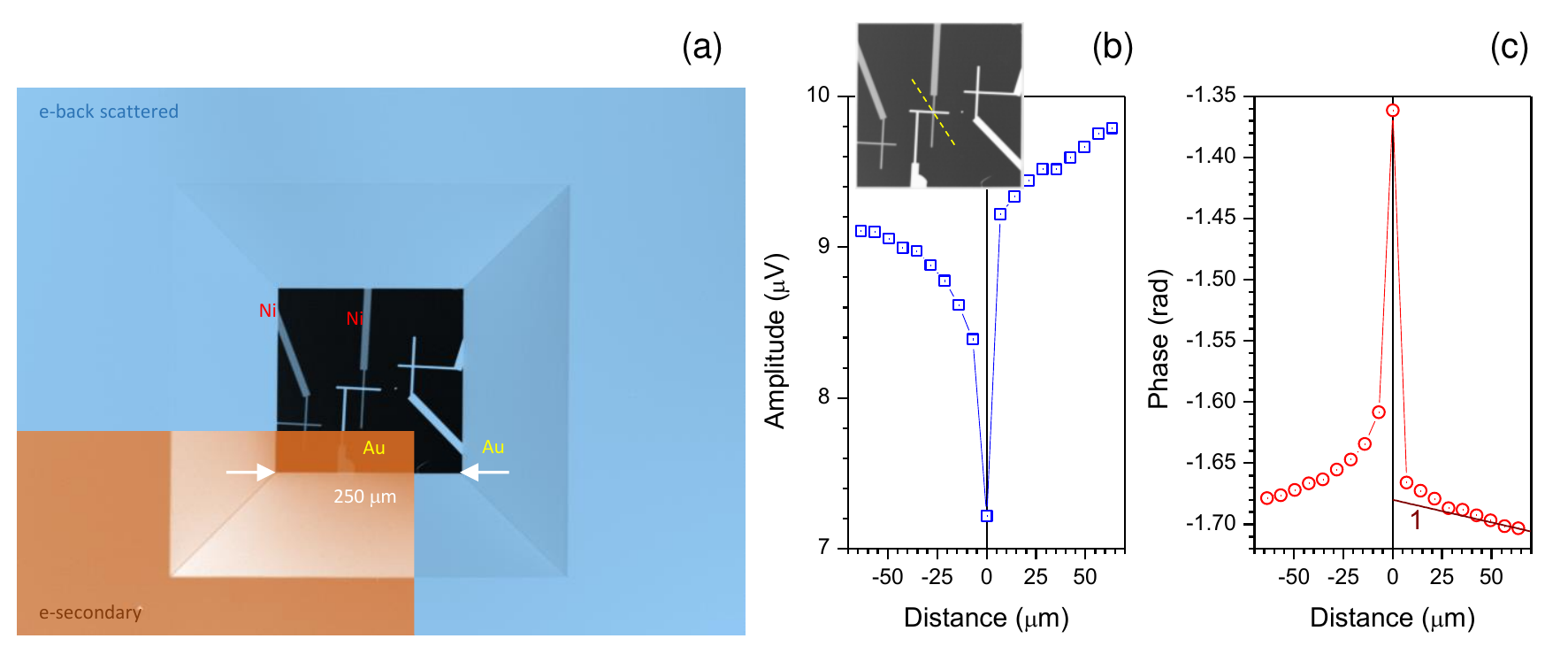}
\caption{Response of micro-thermocouple to back-side electron irradiation. (a) SEM image of thermocouple on 30~nm SiN-membrane by back-scattered (in lens)  and secondary (large angle \blue{scattered}) electrons. (b,c) Measured amplitude and phase response of the Au-Ni thermocouple to a diagonal scan (dashed line in \blue{the} inset in (b)) with $\sim 7~\mu$m steps measured with a lock-in amplifier. Thermocouple voltage was normalised to the transmitted electron current measured by the Faraday cup using an additional lock-in amplifier;  e-beam blanking frequency was 27~Hz. The slope of line \emph{(1)} in (c) corresponds to the best fit by a linear $Phase\sim Const\times d$ dependence, where $d$ is the distance between heat source and measurement point (only valid at large separation\blue{s}); the temperature diffusivity was $a = 0.62\times 10^{-4}$~m$^2$/s or 47.7 times larger than that of SiN~\cite{Grigoropoulos}.}
\label{f-back}
\end{figure}

When the thermocouple response was measured from the front-side  (where metal contacts were deposited), larger $\sim 30~\mu$m steps were used and positions were chosen to avoid direct electron irradiation of the thermocouple (Fig.~\ref{f-thin}). Next, a back-side e-beam irradiation was carried out in small $\sim 7~\mu$m steps with simultaneous detection of back-scattered  (reflection) and transmitted (Faraday cup) electrons during measurements of \blue{the} Au-Ni thermocouple response (Fig.~\ref{f-back}). High transmission of SiN-membrane was confirmed with no reflected electrons measured with a detector at a large scattering angle (sensitive to the secondary electrons). Transmission of the membrane to electrons is also confirmed by a high-contract SEM image (Fig.~\ref{f-back}(a)). Thermocouple signal was normalised to the transmitted Faraday cup signal and the surface of the Au-Au and Ni-Ni junctions were grounded to eliminate possible charging effects. All the device area was covered with stainless steel foil with an only 4-mm-opening for the e-beam exposure. Small voltage detected by thermocouple (Fig.~\ref{f-back}(b)) close to Au-Ni junction is a signature of a changing thermopower since direct electron injection into the junction occur. Thermopower of the free electron gas has a negative sign, hence, a reduction of Seebeck coefficient is expected. 

The phase of a lock-in amplifier signal showed an expected phase delay as the e-beam was more distant from the thermocouple junction (Fig.~\ref{f-back}(c)). The phase delay of a heat wave generated by e-beam $\sim 0.5~\mu$m-diameter heat source at the $f = 27$~Hz is expected to follow $\Delta\theta = -\sqrt{\pi f/a}d - \beta$, where $a$~[m$^2$/s] is temperature diffusivity, $d$ is the distance between the heat source and thermocouple, and $\beta$ is instrument constant~\cite{Morikawa}. Temperature diffusivity of a 600-nm-thick SiN-membrane was measured and $a = 1.3\times 10^{-6}$~m$^2$/s value was determined~\cite{Grigoropoulos} while that of gold is $\sim 1.2\times 10^{-4}$~m$^2$/s and $\sim 0.2\times 10^{-4}$~m$^2$/s for Ni. The linear expression between phase and distance is valid at larger separation between the heat source and the temperature measurement point (see line \emph(1) in (c)); \blue{detailed modeling of thermal transport for the used thermocouple on the membrane is presented in Supplementary material section}. The fit was achieved for a $\sim 47.7$ times larger temperature diffusivity $a = 0.62\times 10^{-4}$~m$^2$/s than that of SiN~\cite{Grigoropoulos}. This value is higher than \blue{a} typical value for Ni and approximately twice lower than that of gold. \blue{The electron beam induced heating is one of the contributions to the detected signal in addition to the charge injection which has a strong impact onto an effective thermopower of the metallic segments of thermocouple.}      
 
 %_________________________Fig. 5
 
\section*{Discussion}

Sensitivity of micro-thermocouple to heating by light and electron beam are demonstrated with the same device. Photo-sensitivity of Ni-Ni junction was observed and was much higher than Au-Au which was used for the reference. This \blue{is} caused by formation of Schottky junction and oxidation of Ni, e.g., optically transparent Ni films sputtered on Si creates a solar cell~\cite{16ijhe11941}. \blue{Future studies and calibration of Ni-NiO, Au-Cr, and rectenna~\cite{3} metal-insulator-metal structures for temperature detection are strongly required. } 

The results of e-beam irradiation of thermocouple from the \blue{back-}side \blue{through the} SiN-membrane showed anomalous behavior of a smaller amplitude (voltage) with irradiation point closer to the thermocouple with a minimum when the e-beam is focused onto the junction (Fig.~\ref{f-back}(b)). During \blue{the} measurements it was observed that \blue{a} longer equilibration time  (minutes)  was necessary for the amplitude and phase to be stable when e-beam was irradiated on the junction. Lower amplitude (voltage) would be equivalent to the smaller temperature for the fixed value of Seebeck coefficient (thermopower). However, injection of electrons into the junction is equivalent to creation \blue{a} more conductive region \blue{with} a higher electron density. The higher the density, the more negative values of Seebeck coefficient are expected as for the free electron gas. The phase signal (Fig.~\ref{f-back}(c)) \blue{at larger separation of the e-beam irradiation from thermocouple} is consistent with the temperature diffusivity of gold rather than SiN-membrane as determined above (Secs.~Results \blue{and Supplement material}). An electron injection into micro-junctions by direct e-beam irradiation is a complex and not well controlled phenomenon which can be investigated with the miniaturised thermocouples \blue{made for this study}. Separation of the pure thermal phenomenon from a dynamic change of thermopower under increased electron density revealed in this study needs further investigation. \blue{The thermal modeling of the fabricated thermocouple on a membrane (Supplement) shows a frequency response and the signal (proportional to temperature) detected by lock-in-amplifier and predicts dissipation of heat dominated by SiN membrane. In experiments (Fig.~\ref{f-back}(b,c)) however, the faster dissipation was observed corroborating contribution of electron injection and dissipation through metallic leads of thermocouple.}      

\section*{Conclusions}

Thermocouple made \blue{of} micrometer-wide Au-Ni electrodes on \blue{a 30-nm-thick} SiN membrane show sensitivity to optical and electron beam excitation and can be used for direct measurement of temperature. Nano-membrane \blue{decreases a} secondary electron generation which excites a considerably larger sub-surface volume with 1-2~$\mu$m cross sections in bulk samples as measured by Au-Ni micro-thermocouple under e-beam exposure~\cite{Chu1}. The miniaturised heat source by \blue{a} tightly focused laser beam or e-beam accompanied with \blue{a} miniaturised thermocouple opens \blue{a new} toolbox for investigation of heat transport at micro- and nano-scales. \blue{For the absolute temperature measurements, a dedicated calibration of sensitivity is required~\cite{Chu1}. In this study, a fast temporal response was demonstrated with thermocouple-on-a-membrane.} 

The demonstrated thermocouple on a nano-membrane can also be used in optical microscopy applications where thermal registration is decoupled from the sample, e.g., cells in a buffer solution on the opposite side of the membrane. Recently, an electron-beam excited fluorescence from a cell on transparent membrane was optically mapped with resolution down to 50~nm~\cite{Kawata} and a direct measurement of the thermal conditions at the nanoscale could further enhance versatility of such technique. In synchrotron radiation experiments, SiN-membranes are common sample support platforms which could have a thermometer function embedded for \emph{in situ} monitoring of the temperature of the sample.  

\section*{Methods}

Membranes of Si$_3$N$_4$ with different thicknesses of 1~$\mu$m and 30~nm (Norcada Ltd.) were used as substrates for fabrication of thermocouples; thermocouples were also made on slide glass for reference. Micro-thermocouples were made by electron beam lithography (EBL) using a simple scanning electron microscope (ACE-7000/EBU, Sanyu Electron Ltd.). Lithography steps started with definition of a 2.5~$\mu$m-wide Au segment of thermocouple in ZEP520A resist. After development, a 5~nm Cr adhesion layer was evaporated followed with deposition of 50~nm of Au. Then, lift-off was performed in developer. Second step exposure of the Ni segment of the thermocouple was made in ZEP520A. Evaporation of 50~nm of Ni followed by the lift-off. The resulting Au-Ni junction had $\sim 6.3~\mu$m$^2$ area, \blue{which is} smaller than a typical CCD pixel.    

Laser-scribed optical mask was made for definition of secondary contact pads. The mask was superimposed with the lithographically defined thermocouple pattern  for evaporation of 10~nm of Cr followed with 90~nm of Au. Electrical bonding was made with \blue{a silver} paste. Ohmic resistance of Au-Ni thermocouple was typically 500~$\Omega$ and 90~$\Omega$ for Au-Au wire junction of the same geometry. Final device is shown in Fig.~\ref{f-samp}(c) on a glass substrate. We used the established calibration constant of 10.1~$\mu$V/K for Au-Ni thermocouple of similar dimensions~\cite{Chu1}. Calibration of a particular thermocouple can be made using Au-Au (or Ni-Ni) junction fabricated at the close proximity on the same substrate and by measuring ohmic resistance. This measurement is then compared with the direct resistance measurement at the known temperature~\cite{Chu1}. \blue{Design of the thermocouple pattern used in this study does not have a resistance heater which is placed equidistantly from the thermocouple and  reference wires (Au-Au, Ni-Ni). Hence, such calibration was not carried out and we relied on the reported sensitivity~\cite{Chu1}. Calibration of thermocouple for determination of the absolute temperature are important when films of several nanometers are used due to strong differences in Seebeck coefficient (thermopower): it changes from -4 to +14 $\mu$V/K when the film of Cr is increasing in thickness from 5 to 10 nm~\cite{5} (the scale is defined with Platinum having the Seebeck coefficient $S = 0$, $S = -15$ (Ni), $S = +6.5$~$\mu$V/K (Au)). The bulk Cr has thermopower $S = 21.8$~$\mu$V/K~\cite{5}. }      

\bibliography{memb,silk1,paper6b}

\section*{Acknowledgements}
JSPS KAKENHI Grant No.16K06768. NATO grant No. SPS-985048. AB is grateful for research visit support by Tokyo Institute of Technology and Swinburne University. Professor Juodkazis was supported by a travel grant form MEXT The Program for Promoting the Enhancement of Research Universities for  his visit to Tokyo Institute of Technology.

\section*{Author contributions statement}  
JM initiated the project, AB together with MR made  thermocouples and carried out their characterisation. SJ drafted the first version of the manuscript. All authors contributed to discussion of results and writing of the manuscript. 

\section*{Additional information}
% To include, in this order: %\textbf{Accession codes} (where applicable); 
No competing financial and non-financial interests. 

% The corresponding author is responsible for submitting a \href{http://www.nature.com/srep/policies/index.html#competing}{competing financial interests statement} on behalf of all authors of the paper. This statement must be included in the submitted article file.

% \begin{table}[ht]
% \centering
% \begin{tabular}{|l|l|l|}
% \hline
% Condition & n & p \\
% \hline
% A & 5 & 0.1 \\
% \hline
% B & 10 & 0.01 \\
% \hline
% \end{tabular}
% \caption{\label{tab:example}Legend (350 words max). Example legend text.}
% \end{table}

\end{document}